\documentclass[12pt]{iopart}

\usepackage{regge}

\newcommand{\mytitle}{The Simplicial Ricci Tensor}
\usepackage
[breaklinks=true]{hyperref}%
\expandafter\ifx\csname package@font\endcsname\relax\else
 \expandafter\expandafter
 \expandafter\usepackage
 \expandafter\expandafter
 \expandafter{\csname package@font\endcsname}%
\fi
\hypersetup{
  pdftitle=\mytitle,
  pdfauthor= {Paul Alsing, Jonathan R. McDonald, Warner A. Miller},
  pdfsubject={General Relativity},
  pdfkeywords = {Regge calculus, Ricci tensor, discrete differential
    geometry}''Ricci Tensor in Regge Calculus
}

\begin{document}

\title{\mytitle}%

\author{ Paul M. Alsing${}^{1}$, Jonathan
  R. McDonald ${}^{1,2}$ \& Warner A. Miller${}^{3}$}

\address{$^{1}$Information Directorate, Air Force Research Laboratory, Rome, New York 13441\\
$^{2}$Insitut f\"{u}r Angewandte Mathematik,
  Friedrich-Schiller-Universit\"at-Jena, 07743  Jena, Germany \\
  $^{3}$Department of Physics, Florida Atlantic University, Boca
  Raton, FL 33431}

\ead{jonathan.mcdonald.ctr@rl.af.mil}

\date{\today}%
\begin{abstract}
   The Ricci tensor ($Ric$) is fundamental to Einstein's geometric theory
  of gravitation. The 3-dimensional $Ric$ of a spacelike surface vanishes
  at the moment of time symmetry for vacuum spacetimes.  The
  4-dimensional $Ric$ is the Einstein tensor for such spacetimes. More
  recently the $Ric$ was used by Hamilton to define a non-linear, diffusive Ricci
  flow (RF) that was fundamental to Perelman's proof of the Poincar\`e
  conjecture.  Analytic applications of RF can be found in many fields
  including general relativity and mathematics. Numerically it has
  been applied broadly to communication networks, medical physics,
  computer design and more.  In this paper, we use Regge calculus (RC)
  to provide the first geometric discretization of the $Ric$.  This result is fundamental for
  higher-dimensional generalizations of discrete RF. We construct
  this tensor on both the simplicial lattice and its dual and prove
  their equivalence.  We show that the $Ric$ is an edge-based weighted
  average of deficit divided by an edge-based  weighted average of  dual
  area  -- an expression similar to the vertex-based weighted average
  of the scalar curvature reported recently.  We use this $Ric$ in a
  third and independent geometric derivation of the RC Einstein tensor
  in arbitrary dimension.
\end{abstract} 
\maketitle

\section{Introduction}\label{sec:intro}

The Ricci curvature tensor ($Ric$) governs the dynamics of geometry in
vacuum general relativity.  It also has been pivotal in the
mathematical classification of manifolds.  It can therefore have a
profound impact on our understanding of geometry and deepen our
insights into classical and quantum gravity.  Hamilton used the
$Ric$ to define a diffusive curvature flow that is referred to as
Ricci flow (RF) \cite{Hamilton:Ricci};
\begin{equation}
  \left( \begin{array}{c} Rate\  of\ change\\  of\  the \
      metric \end{array} \right)  = -2\, Ric.
\end{equation}
This was instrumental in Perelman's proof of Poincar\'{e}'s
conjecture \cite{Perelman:Entropy, Perelman:Surgery, Perelman:Time}.
In addition to its mathematical applications, RF has been applied to a
broad range of problems ranging from medical physics to network
routing, and from face recognition to general relativity and
cosmology.  Many of the applications of RF are for discrete,
unstructured meshes.  Regge calculus (RC) provides a natural discrete
description of Einstein's geometric theory of gravitation
\cite{Regge:1961}.  Here we apply RC to define the $Ric$ in RC for
arbitrary dimensions, so that RF can be extended to higher dimensions.

Evolutions of the $Ric$ have found recent applications in the physics
of spacetime.  RF is expected to be an important tool for the study of
generic black hole solutions of spacetime.  For example, RF provides a
means for a better understanding of quasi-local mass in non-trivial
asymptotically flat spacetimes \cite{Woolgar:08}.  Moreover, it may be
useful for a mathematically rigourous prescription for black hole
boundary conditions in the numerical relativity community
\cite{Yau:PC}.  Similary, RF has also been applied to black-hole
physics as a means for determining the Bekenstein-Hawking entropy
\cite{Samuel:07, Solodukhin:07}.  In cosmology, there has been
increased interest in RF as means for understanding the averaging
problem in $\Lambda$CDM cosmological models \cite{Carfora:84,
  Carfora:08}.

Numerical methods using RF techniques require discrete representations
of the $Ric$ and its corresponding evolution equation.  Current RF
techniques in computational geometry on complex topologies focus on
2-dimensional representations of higher-dimensional data
\cite{Gu:2002, Gu:2003}.  Meanwhile, recent numerical simulations of
relativistic models examined RF on higher dimensional manifolds with
lower complexity topologies \cite{Wiseman:RF:06, Wiseman:RF:2010,
  Holzegel:07, Isenberg:RF:2008}.  Geometric discretizations of the
$Ric$ are needed for numerical simulation of RF on higher dimensional
manifolds with arbitrary topology.  RC is a natural setting for
investigating the $Ric$ and RF due to its piecewise-flat,
coordinate-free construction which naturally captures the Riemannian
curvature on each codimension $2$ hinge, $h$, of the simplicial
lattice.  Here we use this RC Riemann curvature to derive a simplicial
representation of the $Ric$.  This one-form expression is valid in
arbitrary dimension. We start by reviewing some of the principles
related to representation of differential forms in RC and the notation
used in this article in Section~\ref{sec:notation}.  In
Section~\ref{sec:ricci} we develop simplicial $Ric$ on edges of the
simplicial and dual lattices.  In Section~\ref{sec:einstein} we use
our expression of the simplicial $Ric$ to provide a third and
independent geometric derivation of the RC Einstein tensor in
arbitrary dimension.  In particular, we utilize the simplicial $Ric$
and scalar curvature to explicity construct the Einstein tensor as the
trace-reversed $Ric$.

\section{Dual Lattices and Discrete Differential Forms} \label{sec:notation}

Geometric discretizations \cite{ Gu:2002, Gu:2003, Whitney:book,
  Bossavit:1991, Bossavit:1998, Arnold:2006} are generally
characterized by association of tensors with lattice elements of a
discrete manifold.  Tensors decomposed into the {\em space of values}
and {\em tangent space} components become weighted distributions over
the skeleton of the discrete manifold and obtain their geometric
properties from the skeleton itself.  Differential quantities in the
lattice are formulated such that point-wise evaluation gives way to
averaged evaluation over an integral domain.  Tensors thus become
integrated measures on the discrete manifold and their associated
scalar weights may be intepreted as densities assigned to a lattice
element.  This integrated representation of tensors over lattice
elements is a form of discrete exterior calculus or discrete
differential forms (DDF) in which one explicitly discretizes the
tangent space values of a differential form.

The simplicial lattice in RC provides one set of differential forms
onto which a tensor may be projected.  The simplicial $d$-volumes of a
$d$-dimensional manifold provide an anchor -- the tangent space-- for
the differential forms.  However, to incorporate dual forms we require a
lattice structure obtained by some duality relation with the simplicial
skeleton, i.e. the dual lattice.  We will often use the more generic
phrasing of dual lattice to refer to the circumcentric dual lattice.
The circumcentric dual lattice is the unique lattice defined by
connecting the circumcenter of a $d$-simplex to the circumcenters of
each neighboring $d$-simplex.  This lattice is of special interest
since it creates a pair-wise orthogonality between elements of the
dual lattice, i.e. for each $k$-element of the simplicial lattice
there exists a ($d-k$)-element in the circumcentric dual.  Moreover,
if we constrain the simplicial lattice to be a Delaunay lattice
\cite{Okabe:book}, then the circumcentric dual is identified as a
Voronoi lattice.  In this particular case, the $d$-dimensional Voronoi
cells are uniquely determined by the set of all points closest to a
given simplicial vertex than to any other simplicial vertex.
Likewise, a general ($d-k$)-Voronoi element is the set of all points
in the codimension-$k$ hyperplane closest to a $k$-simplex than to any
other $k$-simplex in the simplicial lattice.  Thus, a $d$-volume
constructed from the simplicial element and its Voronoi dual has a
natural interpretation as the local, compact integral measures on the
simplicial lattice. (See~\ref{App:Volumes} for more details.)

Some of the notation used in this article will denote elements of the
simplicial (dual) lattice, volumes in the lattices, or measures of
curvature.  In particular, we will distinguish between the simplicial
and dual lattices using Latin and Greek lettering. The Latin letters
$\vs$, $\ls$, and $\tris$ will label simplicial elements of dimension 0, 1,
and 2, respectively.  Arbitrary $k$-simplexes are labeled by
$\polys{k}$.  Meanwhile, the elements of the dual lattice are labeled by
the Greek letter counterparts $\vd$, $\ld$, $\trid$, and
$\polyd{k}$.  We will also be using the notation $\Delta V_{a}$ to
denote the $d$-volume associated with the element $a$.  For an
edge $\ls$ of the simplicial lattice on a 3-dimensional lattice, the
label $\Delta V_{\ls}$ represents a 3-volume associated with $\ls$.  The
label $\resvol{a}{b}$ denotes the $d$-volume associated with $a$ and
restricted to the element $b$.  This restriction can be formulated as
taking the intersection of the individual $d$-volumes from $a$ and
$b$.  Extending this notation to arbitrary restrictions, we can write
$\Delta V_{a_{1} a_{2} \cdots a_{k}}$ as the restriction of the volume
$\Delta V_{a_{1}}$ to all of the elements $a_{2}, \ldots, a_{k}$.
Indeed, one can convince oneself of this notation by considering the
case of the simplicial manifold restricted to a given element of
either lattice.  In this case, the entire manifold contains the $d$-volume
of every lattice element, so $\Delta V_{a}$ can be seen to be the
restriction of the simplicial manifold to the element $a$.

These notations, and others, are summarized below:

\begin{tabular} {r c l}
$\vd$ & -- & Dual vertex \\
$\ld$ &--&  Dual edge \\
$\trid$, $h^{*}$ &--& Dual polygon \\
$\polyd{k}$ &--& Dual polytope of dimension $k$\\
$\vs$ &--&  Simplicial vertex\\
$\ls$ &--& Simplicial edge\\
$\tris$ &--& Triangle on simplicial skeleton \\
$\polys{k}$& --& $k$-simplex\\
$h$ &--& Simplicial hinge \\
$St(a)$ &--& Star of a lattice element $a$,
i.e. $\bigcup_{s^{(k)}\supset a } s^{(k)}$ for the simplicial lattice\\
$A_{h}$, $|h|$ &--& Volume of $h$ \\
$A_{h}^{*}$, $|h^{*}|$ &--& Area of $h^{*}$ \\
$|s^{(k)}|$ $\left( |\sigma^{(k)}|\right)$ &--& volume of $s^{(k)}$
$\left( \sigma^{(k)}\right)$ \\
$\theta_{h\ls}$ &--& Angle opposite of edge $\ls$ on a hinge $h$ in 4
dimensions \\
$\epsilon_{h}$ &--& deficit angle associated with a hinge \\
$R_{h}$& --& Riemann Tensor projected on a hinge \\
$R_{\lambda}$&--& $Ric$ projected on a dual edge, $\ld$\\
$R_{\ls}$&--& $Ric$ projected on a simplicial edge, $\ls$ \\
$R_{\nu}$& --& Ricci scalar at a dual vertex, $\vd$\\
$R_{v}$ &--& Ricci scalar at a simplicial vertex, $\vs$\\
$A_{h\ls}$ &--& Volume of hinge restricted to $\ls$\\
$A_{h\ld}^{*}$& --&  Area of dual to a hinge restricted to $\ld$\\
$\norm{a}_{b}$ &--& Volume of $a$ restricted to $b$, i.e. the
norm of $a\bigcap b$\\
$\norm{a}_{b_{1} \cdots b_{m}}$ &--& Volume of $a$ restricted to all $b_{i}$, i.e. the
norm of $a\bigcap b_{1}\bigcap \cdots \bigcap b_{m} $\\
$\Delta V_{a}$&-- & $d$-volume associated with the element
  (either dual or simplicial) $a$. \\
$\resvol{a}{b}$ &--&  $d$-volume of $a$ restricted to $b$ \\
$\left< \alpha^{(k)}, s^{(k)}\right>$ & --&  Local projection or metric
  inner-product of two $k$-forms, \\
$\left( \alpha^{(k)}, \beta^{(k)}\right)$ &--& Standard $L_{2}$
  inner-product on two simplicial (dual) $k$-forms\\
${\overline{\left\langle C_{a} \right\rangle}_{b}}$& --  &
 Volume-weighted average of the $C_{a}$'s hinging on the
element $b$, $\frac{ \sum_{a: b\in a}  C_{a} \resvol{a}{b}   }{\sum_{a: b\in a}
  \resvol{a}{b}}$\\
$\left<
  C_{a}\right>_{b}$ &--& Area-weighted average of the $C_{a}$'s hinging on
the element $b$,  $a\frac{\sum_{a: b\in a} C_{a}A_{ab}}{\sum_{a: b
    \in a} A_{ab}}$\\
$\bar{C}_{a}|_{b}$ & -- & Arithmetic mean of the $C_{a}$'s hinging
on $b$ \\
\end{tabular}
 \vspace{3ex}

\section{Discretizing the Ricci Tensor} \label{sec:ricci}

Here we  construct a geometric
representation of the $Ric$ on a piecewise-flat simplicial geometry.   The geometric
discretization we use  is based on discrete differential forms (DDF) in which
the (dual) simplicial lattice is used as the (co-)chain complex for
embedding continuum forms in the discrete manifold.  It has been
found that such discretizations preserve the geometric properties
of the tensors and can be useful for solving differential equations
for tensor fields on geometries with complex topology
\cite{Bossavit:1998, Desbrun:DisForms, Arnold:2006}.  

Piecewise-flat geometries are characterized by curvature distributions
concentrated at each codimension 2 hinge, $h$, on the simplicial
manifold, ${\cal S}$.  The curvature on a given hinge $h$ is a
conical singularity with deficit angle $\epsilon_{h}$.  We have shown
that standard RC is consistent with distributing this curvature evenly
over the polyhedron, $h^*$, (with area $A_{h}^{*}$) dual to hinge $h$.
It admits a natural interpretation as the sole independent component
of the Riemann curvature tensor in the $d$-volume associated with the
hinge \cite{Miller:HARC}.  From this local representation of curvature
distributed over a hinge one can explicitly and geometrically define
the Einstein tensor in 4-dimensions \cite{Miller:BBP} and a
vertex-based scalar curvature \cite{McDonald:RCscalcurv}.  The
Einstein tensor encodes the geometrodynamics of General Relativity
through the Einstein equations.  The scalar curvature provides a
point-wise average of curvature that an observer can set out to
measure.  However, these curvature measures are insufficient to
examine geometric flows where the $Ric$ plays the predominent role.
When discretizing evolution processes that can be reformulated as an
evolution of the $Ric$ itself, e.g. RF, we seek to first represent the
$Ric$ directly in the geometry, then develop the evolution equations
for the new representation.  We provide two equivalent derivations of
the $Ric$. First, we start with the continuum construction and apply
it directly to discrete curvature forms.  Second, we derive an
equivalent expression directly from the action principle of RC.

\subsection{Derivation of the $Ric$ from the Continuum using  Discrete Curvature Forms}
\label{sec:RICDCF}

In the continuum the $Ric$ is defined as the first contraction
of the Riemann curvature tensor;
\begin{equation}
R^{a}_{\phantom{a}b} = R^{ac}_{\phantom{acd}bc}.
\end{equation}
As a bivector-valued two-form the curvature tensor takes in a
bivector for the loop of parallel transport and outputs a
bivector characterizing the change in a vector transported around the
loop;
\begin{equation}
{\bf R} =\frac{1}{4} e_{a} \wedge e_{b}\; R^{ab}_{\phantom{ab}cd} dx^{c}\;  \wedge dx^{d}
\end{equation}
where $\{e_{a}\}$ are the basis tangent vectors dual to the basis
one-forms $\{dx^{c}\}$.

 In RC, curvature is given exclusively
by the sectional curvature, $K$, associated with a codimension 2
hinge.  On a hinge, the sectional curvature is given by
\begin{equation}
K = \frac{\epsilon_{h}}{A_{h}^*}
\end{equation}
which is just the ratio of angle rotated (the deficit angle
$epsilon_{h}$) to area traversed ($A_{h}^{*}$) by the loop of parallel
transport.  The sectional curvature is the double projection of the
Riemann tensor onto a given plane \cite{Kobayashi:diffgeom};
\begin{equation}
K = {\bf R}(e_{a}, e_{b}, e_{a}, e_{b})
\end{equation}
where $e_{a}$ and $e_{b}$ are an orthonormal basis for the plane.
Hence the Riemann curvature tensor on a hinge is proportional to the
sectional curvature of the polygonal dual, $h^{*}$, to the hinge;
\begin{equation}
 {\cal R}_{h} =R(h^{*}_{ab}, h^{*, ab}) = d(d-1) \frac{\epsilon_{h}}{A_{h}^{*}}.
\end{equation}
For this reason, one can generally denote the Riemann tensor for a
hinge as $R^{h^{*}}_{\phantom{h^{*}}h^{*}}$.  We will, in general,
only keep track of the two-form components and write $R_{h^{*}}=
R_{h}$, where the equality is a result of the duality. Taking the
trace of the Riemann tensor requires summation over the curvature
associated with loops spanned, in part, by a given one-form $e^{b}$.
This summation of loops hinging on a given one-form reduces the
curvature two-form to a one-form doubly-projected on $e^{b}$.

\begin{figure}[ct]
\centering
\includegraphics[height=2.5in]{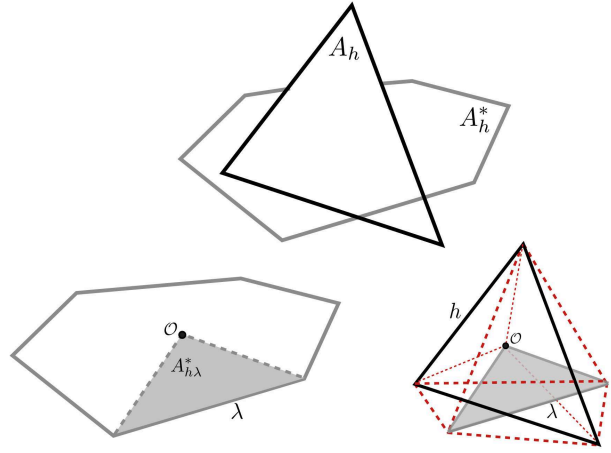}
\caption{{\bf Restricting the Hinge Volume to a Dual Edge:} Here we
  explicitly show the decomposition of the d-volume of a hinge $h$ (in
  $d=4$) and its restriction to a dual edge $\ld$.  ({\it Top})  Here we
  show the orthogonal decomposition of the $d$-volume into the area
  of a hinge, $A_{h}$, and the area of the dual polygon to a hinge,
  $A_{h}^{*}$.  Struts (not-shown) connecting each vertex of $h^{*}$
  to each vertex of $h$ complete the boundary of the domain spanned by
  $h$ and $h^{*}$.  ({\it Bottom-left}) We focus attention on the dual polygon
  $h^{*}$ and have shown (shaded) the restriction to the dual edge
  $\ld$.  This restricted area is the 2-simplex constructed from
  the endpoints of $\ld$ and the circumcenter of the hinge $h$.
  ({\it Bottom-right})  By connecting the vertexes of the restricted area of $h^{*}$,
  $A_{h\ld}^{*}$, to each of the vertexes of the hinge $h$, we
  obtain a new $d$-volume, $\Delta V_{h^{*}\ld} = \Delta
  V_{h\ld}$.  The thick red (dashed) lines are struts connecting
  vertexes on the boundary of $\Delta V_{h^{*}\ld}$   The  struts
  connecting the circumcenter ${\cal O}$ of $h$ to the vertexes of $h$
  (thin dashed, red) do not contribute the boundary of $\Delta
  V_{h\ld}$ and can be routinely dropped from the construction.}  
\label{Fig1}
\end{figure}

RC is at its heart a weak variational formulation of General
Relativity.  This is easily seen since the geometric content of RC is
encoded not through pointwise defined tensors, but tensors distributed
over elements of the lattice.  Indeed, the Regge equations are
integral equations and given by the Einstein tensor integrated over
the associated 4-volume.  Hence, we evaluate the discrete $Ric$ as an
integrated quantity on the simplicial manifold.  Locally, the $Ric$
becomes a one-form projected on the dual edges of the lattice and
integrated over the $d$-dimensional domain, $\Delta V_{\ld}$,
associated with the dual edge, $\ld$.  To take the trace of the
Riemann tensor, one must sum over the independent directions
orthogonal to a dual-edge $\ld$.  In general, one will sum over
all independent two-forms $\ld \wedge e^{a}$.  However, when
$e^{a}$ lies in the plane of a hinge $h$, there is no curvature
associated with such a loop of parallel transport.  Therefore, the
Ricci one-form on $\ld$ is dependent only on the the polyhedral
2-faces, $h^{*}$, hinging on a given dual edge, $\ld$;
\begin{equation}
  R_{\ld}\Delta V_{\ld} = \sum_{h^{*}: \ld \in h^{*}} R_{h^*}
  \resvol{h^{*}}{\ld}.
\end{equation} 
We have introduced the volume $\resvol{h^{*}}{\ld}$ (Figure~\ref{Fig1}) which is a
restriction of the $d$-volume for $h^{*}$ to
the dual edge $\ld$--the intersection of the $d$-volumes
associated with $h^{*}$ and $\ld$.    
This is the discrete equivalent of decomposing a domain and
integrating over distinct representations on the subdomains;
\begin{equation}
  \int_{\Omega} \alpha  = \sum_{i} \int _{\Omega_{i}}
  \alpha'(\Omega_{i}).
     \quad\quad 
\end{equation}
Here the $\Omega_{i}$ form a non-overlapping domain decomposition of
$\Omega$.   Using the Voronoi-Delaunay orthogonal decomposition of
volumes and the RC definition of curvature on a hinge, $R_{h} = {d(d-1)}
\frac{\epsilon_{h}}{A_{h}^{*}}$,  we obtain an explicit expression for the
integrated Ricci one-form on a dual edge; 
\begin{eqnarray}\label{eq:RicciDualForm}
 R_{\ld} \Delta V_{\ld} &=& \sum_{h^{*}:\; \ld \in h^{*}} {d(d-1)}
  \frac{\epsilon_{h}}{A_{h}^{*}} 
    \frac{1}{\binom{d}{2}} A_{h} A_{h\ld}^{*} \nonumber \\
 &=& \sum_{h^{*}: \; \ld \in h^{*}} 2 \epsilon_{h}A_{h} \frac{A_{h\ld}^{*}}{A_{h}^{*}}.
\end{eqnarray}
We have decomposed the restricted $d$-volume,
(see~\ref{App:Volumes}), into the Voronoi and Delaunay components and
restricted the Voronoi area, $A_{h}^{*}$, to the dual edge,
$\ld$, denoted as $A_{h\ld}^{*}$.  This restricted area is the
set of all points in $A_{h}^{*}$ closer to $\ld$ than to any other
dual edge $\ld'$ in the skeleton of $h^{*}$.  Dividing by
the intergal domain, we obtain
\begin{equation}
R_{\ld} = \frac{   \sum_{h^{*} \; : \ld \in h^{*}} {d(d-1)}
  \frac{\epsilon_{h}}{A_{h}^{*}} A_{h\ld}^{*}A_{h}}{   \sum_{h^{*}
    \; : \lambda \in h^{*}}  A_{h\ld}^{*} A_{h}} = \frac{   \sum_{h^{*} \; : \ld \in h^{*}} R_{h}\resvol{h}{\ld}}{   \sum_{h^{*}
    \; : \lambda \in h^{*}} \resvol{h}{\ld}} . 
\end{equation}
Defining a volume-weighted average as 
\begin{equation*}
{\overline{\left\langle C_{a} \right\rangle}_{b}}=
\frac{ \sum_{a: b\in a}  C_{a} \resvol{a}{b}   }{\sum_{a: b\in a} \resvol{a}{b}   },
\end{equation*}
 the Ricci one-form in the
dual lattice becomes
\begin{equation}
R_{\ld} = \overline{\left\langle R_{h}\right\rangle}_{\ld}.   
\end{equation}
This is an explicit expression of the $Ric$ in the dual lattice
as a weighted average of curvatures meeting on the dual lattice one-form
$\lambda$. 

In RC, it is customary for measures of curvature to be associated with
elements in the simplicial lattice. This more readily allows for
evolution equations in terms of the degrees of freedom, the edge
lengths of the simplexes $\{\ls\}$.  For applications of the $Ric$, such
as for RF, this is particularly important since a straightforward weak
evolution equation for the edge lengths (synonomous with the
components of metric) will require an integration of the $Ric$ over
the $d$-volume associated with an edge, i.e. the integrated Ricci
one-form at a given $\ls$.  We thus seek to re-express the Ricci
one-form on the simplicial skeleton.  Taking the dual of the above
expression gives us a Ricci three-form on the simplicial lattice.
However, it is beneficial to write an explicit expression for the
Ricci one-form on simplicial edges.  We can transform the above
expression into a edge-based expression in the simplicial skeleton via
a lowering (raising) operator which transforms $r$-forms in the dual
(simplicial) lattice to $r$-forms in the simplicial (dual) lattice
(see~\ref{App:DisFormOps}).  We first rewrite the association of the
$Ric$ on a dual edge by restricting the domain to that closest to a
simplicial edge, $\ls$.  This is the result of the projection of the
dual edge $Ric$ onto the domain of the edge, $\ls$;
\begin{eqnarray}
R_{\ld}\resvol{\ld}{\ls} &=& R_{\ld}\Delta V_{\ld}
\frac{\resvol{\ld}{\ls} }{\Delta V_{\ld}}  \nonumber \\
& =& \sum_{h^{*}: \ld \in h^{*}} R_{h^*} \resvol{h^{*}}{\ld\, \ls}.
\end{eqnarray}

For $d>2$ this newly projected volume can be decomposed as before, except now  we
must restrict the hinge area to that which is closest to $\ls$.
Suitably rearranging the terms in the sums gives
\begin{eqnarray}
R_{\ls} \Delta V_{\ls} &=& \sum_{ \ld \in \ls^{*}}  R_{\ld}
\resvol{\ld}{\ls}  \label{eq:RicciSimp1} \\
&=& \sum _{\ld\in \ls^{*}} \;\; \sum_{h^{*}:\; \ld \in h^{*}} 2 \epsilon_{h}A_{h\ls}
\frac{A_{h\ld}^{*}}{A_{h}^{*}}  \;\;\;\;\;\;\;\;\; {\rm (for}\;
d>2{\rm )}\nonumber \\
&=& \sum_{h: \; \ls\in h} \frac{2\epsilon_{h} A_{h\ls}}{A_{h}^{*}} \sum_{\ld \in h^{*}}  
{A_{h\ld}^{*}}  \nonumber \\
&=& 2 \left<\epsilon_{h}\right>_{\ls}A_{h}
\end{eqnarray}
where we have defined the edge-based area-weighted average as $$\left<
  C_{h}\right>_{\ls}~=~\frac{\sum_{h: \ls\in h} C_{h}A_{h\ls}}{\sum_{h: \ls
    \in h} A_{h\ls}}.$$ It is key to note here that swapping the
summations is allowed given that the Voronoi-Delaunay decomposition of
the volumes determines a tiling of the manifold without overlap.  This
will generally be true for arbitrary triangulations with
circumcentric duals as long as volume orientation is also carried
over in the calculation.  Again, dividing by the integral volume, we
get an explicit expression for the $Ric$ weighting on an edge
of the simplicial lattice;
\begin{equation}
  R_{\ls} = {d(d-1)}
  \frac{\left<\epsilon_{h}\right>_{\ls}}{\left<A_{h^{*}}\right>_{\ls}}
  \;\;\;\;\; {\rm (for}\;   d>2{\rm )}.
\end{equation}

\begin{figure}[ct]
\centering
\includegraphics[height=2.5in]{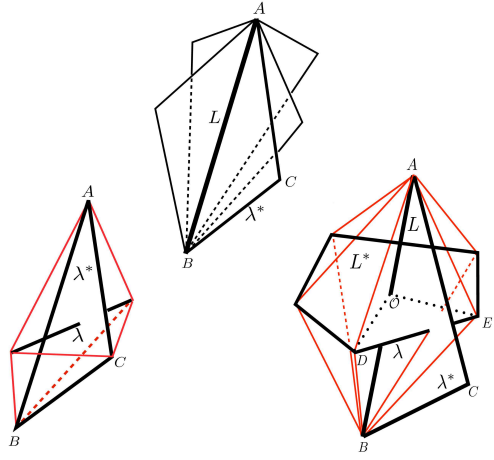}
\caption{{\bf Volumes for the $Ric$ on a Simplicial Hinge:}
  Here we use the case of $d=3$ as a concrete example of the
  construction of the simplicial $Ric$ from the dual edge-based
  $Ric$.  ({\it Top}) A simplicial edge $\ls$ is shown with all
  triangles $t=\ld^{*}$ hinging on $\ls$.  The notation of
  $t=\ld^{*}$ indicates that to each triangle containing $\ls$,
  there is an edge $\ld$ of the dual lattice orthogonal and dual
  to $t$.  ({\it Bottom-left}) The 3-volume for a given $\ld$ is
  depicted here.  In general, only a portion of this volume will
  overlap with the 3-volume associated with $\ls$.  To construct the
  $Ric$ for $\ld$, the integral volumes used must coincide, so we
  take the restriction of $\Delta V_{\ld}$ to $\ls$, $\Delta
  V_{\ld \ls}$.  ({\it Bottom-right}) The 3-volume $\Delta V_{\ls}$ is
  shown and we indicate the part of $\Delta V_{\ls}$ corresponding to
  $\Delta V_{\ld \ls}$ as the volume spanned by the vertexes $[
  ABDE{\cal O}]$.  Since $\ls^{*} = h^{*}$, summing over all $\ld$
  contained in $h^{*}$ carries us around the loop orthogonal to $\ls$.
  In the restriction of $\Delta V_{\ld}$ to $\ls$, the only
  contribution with non-trivial restricted volume is $h^{*} = \ls^{*}$
  for the given $\ls$.  Hence, substituting the expression for
  $R_{\ld}$ into Eq.~\ref{eq:RicciSimp1} and summing over all
  $\lambda \in \ls^{*}=h^{*}$, we obtain the Regge curvature on an
  edge/hinge in $d=3$.  Hence we have $R_{\ls} \Delta V_{\ls}= R_{h}
  \Delta V_{h}$ as expected. }
\label{Fig2}
\end{figure}

For the special case of $d=3$, the Riemann tensor is proportional to
the $Ric$, i.e. all curvature content is encoded
directly in the Ricci tensor.  In Figure~\ref{Fig2} we look at the
$Ric$ on a simplicial edge in $d=3$ and illustrate the volumes
associated with the construction. 

We now turn to the special case of $d=2$.  The duality between $\ld$ and
$\ls$ is such that
\begin{eqnarray}
R_{\ls} \Delta V_{\ls} &=& \sum_{ \ld \in \ls^{*}}  R_{\ld}
\resvol{\ld}{\ls}   \nonumber \\
&=& \sum_{\ld} R_{\ld} \Delta V_{\ld} \delta_{\ld,\ls^{*}} = R_{\ld}\Delta V_{\ld}.
\end{eqnarray}
Again, the duality can be used to show $R_{\ls}=R_{\ld}$.   Using
the expression for $R_{\ld} = \overline{\left\langle R_{h=\vs}\right\rangle}_{\ld}$
we have
\begin{eqnarray}
R_{\ls} &=& \frac{\sum_{h=\vs \ni \ls}R_{h}
  A^{*}_{h\ls}}  {\sum_{h=\vs \ni \ls} A^{*}_{h\ls}} \nonumber \\
&=&  \frac{\sum_{h=\vs\ni \ls}R_{h}
 \frac{1}{4} \ls \times \ls^{*}}  { \frac{1}{2}\ls \times
 \ls^{*}} \nonumber \\
&=&\bar{R}_{h}|_{\ls}
\end{eqnarray}
where $\bar{R}_{h}|_{\ls}$ is the arithmetic average of the curvature
evaluated at the endpoints of $\ls$.  We have used the normalization of
${\rm Vol}(h=\vs) =1$ in the first equality and the relation $A_{h\ls}=
\frac{1}{4} \ls \times \ls^{*} = A_{h'\ls}$ for both endpoints (hinges), $h$
and $h'$, on $\ls$ in the second equality.  This is a discrete
expression showing explicitly that the Ricci one-form is determined
solely by the scalar curvature (on vertexes) in 2-dimensions.

\subsection{Derivation of the $Ric$ from the RC Action Principle}
\label{sec:RICAP}

These expressions can also be derived from the Regge action principle
in a similar way to the authors' previous construction of the scalar
curvature invariant in RC \cite{McDonald:RCscalcurv}.  Since the
curvature is locally proportional to the sectional curvature, we
obtain a simple relation between the action using the curvature
two-form and the canonical Einstein-Hilbert action;
\begin{equation}
I =\frac{1}{\kappa} \sum_{h}({\bf R}, {\bf h}^{*})  = \frac{1}{\kappa}
\sum_{h} d(d-1) K_{h} \Delta V_{h} = \frac{2}{\kappa} \sum_{h}
\epsilon_{h} A_{h} = I_{\rm Regge}
\end{equation} 
where ${\bf h}^{*}$ is the two-form for the dual loop to a hinge and $K_{h}
= \frac{\epsilon_{h}}{A_{h}^{*}}$ is the sectional curvature.  The
factor of $d(d-1)$ comes about from contracting the curvature two-form
with the dual polygon two-form which gives equal contributions from
all non-zero components.  Using duality, we can also change the first
expression to a hinge-based, instead of a dual polygon, expression;
\begin{equation}
I =\frac{1}{\kappa} \sum_{h}(\star {\bf R}, {\bf h}). 
\end{equation}
Now tracing over directions orthogonal to edges and summing over the
edges we get
\begin{equation}
I=\frac{1}{\kappa} \sum_{h} \sum_{\ls \in h} (d-1) \left( R_{h\ls}, \ls\right)
\end{equation}
where $R_{h\ls}$ is the $Ric$ on hinge $h$ directed along $\ls$.
Contracting the $Ric$ with its associate one-form gives an
additional factor of $d$ such that we obtain
\begin{equation}
I =\frac{1}{\kappa} \sum_{h} \sum_{\ls \in h} d(d-1) K_{h} \resvol{\ls}{h}.
\end{equation}
To get the action in terms of the Ricci one-form we 
decompose the integral measures and rearrange the summations;
\begin{eqnarray}
  I  & = &\sum_{h} d(d-1)\frac{\epsilon_{h}}{A_{h}^{*}}
 \underbrace{\frac{1}{ \binom{d}{2}}\sum_{\ls \in h} A_{h}^{*}
   A_{h\ls}}_{\resvol{\ls}{h}}    \nonumber \\
&=&\sum_{h} \sum_{\ls \in h}  d(d-1)\frac{\epsilon_{h}}{A_{h}^{*}}
  \frac{1}{\binom{d}{2}}A_{h}^{*}  A_{h\ls}  \nonumber \\
  &=& \sum_{\ls} \sum_{h:\; \ls\in h} d(d-1)\frac{1}{\binom{d}{2}}
  \frac{\epsilon_{h}}{A_{h}^{*}}  A_{h}^{*} A_{h\ls}  \nonumber \\
  &= &\sum_{\ls} R_{\ls} \Delta V_{\ls}.
\end{eqnarray}

Using the equality of the individual terms in the sum over edges, we
get an expression for the curvature on an edge of the simplicial
lattice;
\begin{equation}
R_{\ls} \Delta V_{\ls} =  \sum_{h:\; \ls\in h} d(d-1)\frac{1}{\binom{d}{2}}
\frac{\epsilon_{h}}{A_{h}^{*}}  A_{h}^{*} A_{h \ls}. 
\end{equation}
We have absorbed the combinatoric factor of $d(d-1)$ into the
definition of $R_{\ls}$ as we will do in general.  This helps keep in
mind that the expression for $R_{\ls}$ is a scalar weight on the edge
element.  Formally, these scalar weights are part of a integrated
quantity and are not necessarily assigned to a point on the lattice,
but rather across the domain of integration associated with the given
element.  Hence, the curvature forms used in RC are to be understood
as $R_{h}\Delta V_{h}$, $R_{\ls}\Delta V_{\ls}$, and $R_{v}\Delta V_{v}$
for the Riemann, Ricci and scalar curvature, respectively.

We can raise the simplicial Ricci one-form to obtain the dual Ricci
one-form.  To do so, we first restrict the integrative domain to the
volume closest to the dual edge;
\begin{equation}
R_{\ls} \resvol{\ls}{\ld} =  \sum_{h:\; \ls\in h} \frac{d(d-1)}{\binom{d}{2}}
\frac{\epsilon_{h}}{A_{h}^{*}} A_{h\ld}^{*} A_{h \ls}.
\end{equation}
We define the raising (lowering) operation applied to the simplicial
$Ric$ by summing over all integrated $R_\ls$ for which $\ld\in \ls^{*}$.  The
restriction of the domain above is necessary to ensure that lowering
(raising) this expression gives a quantity integrated over the
appropriate $d$-volume.  Doing so we obtain
\begin{eqnarray}
R_{\ld}\Delta V_{\ld}  &= \sum_{\ls:\,\, \ld \in \ls^{*}} R_{\ls}
\resvol{\ls}{\ld} \nonumber \\ 
&= \sum_{\ls:\,\, l \in \ls^{*}} \frac{d(d-1)}{\binom{d}{2}}
\sum_{h:\;{\ls}\in h}\frac{\epsilon_{h}}{A_{h}^{*}}   A_{h\ld}^{*} A_{h\ls}
\nonumber \\ 
&= \sum_{h:\; \ld \in h^{*}}  \frac{d(d-1)}{\binom{d}{2}}
\sum_{\ls\in h}\frac{\epsilon_{h}}{A_{h}^{*}}   A_{h\ld}^{*}  A_{h \ls}
\nonumber \\ 
&  = \sum_{h:\; \ld \in h^{*}}  \frac{d(d-1)}{\binom{d}{2}}
\frac{\epsilon_{h}}{A_{h}^{*}}  A_{h\ld}^{*}  A_{h}   \nonumber \\  
&=  \sum_{h:\; \ld \in h^{*}}  2\frac{\epsilon_{h}}{A_{h}^{*}}  A_{h\ld}^{*}  A_{h}.
\end{eqnarray}

Comparing  with Eq.~(\ref{eq:RicciDualForm}) shows exact
agreement.  The independence of the local and global derivations shown
here are indicative of the decomposition of the lattice into elements
with compact support.   Therefore, the global derivation in terms of
the action becomes just an additional sum over the local terms defined
over the domains of compact, local support.   This highlights the
reason that RC as a weak variational principle reduces to locally
simple characterizations of the manifold geometry. 

\section{The Canonical Einstein Tensor} \label{sec:einstein}
  
In previous work, the Cartan moment-of-rotation trivector view was
used to derive the embedding of the Einstein tensor in RC
\cite{Miller:BBP, McDonald:BBP}.  Here we present an alternative
derivation using the more familiar definition of the Einstein tensor;
\begin{equation} \label{eq:EinsteinTensor}
G_{\mu\nu}  = R_{\mu\nu} - \frac{1}{2} g_{\mu\nu} R
\end{equation} 
which can be rewritten as an Einstein one-form
\begin{equation}
G_{a} \equiv G_{\mu\nu}e^{\nu}_{a}= R_{a} - \frac{1}{2}e_{a} R.
\end{equation}
Using the simplicial $Ric$ and the previously derived scalar
curvature \cite{McDonald:RCscalcurv}, we have all the necessary tools to provide a direct
reconstruction of the Einstein tensor on an edge. 

The isomorphism between forms on the dual and forms on the simplicial
lattice allows us the freedom to define curvature forms on either
lattice.  However, we should start off on a sound geometric footing by
following the projection of the continuum object onto the lattice
structure.  Eq.~(\ref{eq:EinsteinTensor}) identifies the quantitative
construction of the Einstein tensor, but does not indicate the
geometric character of the Einstein one-form.  However, it is known
that the Einstein tensor is the double-dual of the Riemann curvature
tensor \cite{Synge:GR};
\begin{equation}
  G_{i}^{\phantom{i}j}
  =\left({}^{*}R^{*}\right)_{im}^{\phantom{im}jm}=\frac{1}{4}
  \epsilon_{mnil}R^{mn}_{\phantom{mn}ab} \epsilon^{ablj}.
\label{eq:DDRiemann}
\end{equation}
The Hodge duals transform the two-form components on the dual lattice
to forms on the simplicial lattice.  The trace over the second and
third indices reduce the two-form to a one-form.   Hence, the Einstein
tensor is a one-form on edges of the simplicial lattice.  Equivalently,
in 4-d the Einstein one-form is the dual of the moment of rotation
3-form projected on the 3-volume dual to an edge \cite{Miller:BBP,
  McDonald:BBP}.  We take the natural embedding for the Einstein
one-form in RC to be on the simplicial 1-skeleton.  One
could easily construct a dual lattice Einstein one-form, though we see
no particular benefit.

We must also be careful in how we introduce the vertex-based scalar
curvature, $R_{v}$, in the edge-based representation.  This is most directly
accomplished by projecting the integrated scalar curvature at a vertex
onto the $d$-volume associated with an edge;
\begin{equation}
e_{a}R \longrightarrow R_{\vs} \Delta V_{\vs \ls}
\end{equation}
This contributes non-trivially only when the vertex $\vs$ is a vertex of
$\ls$.  Moreover, since the scalar curvature is decomposed into volumes
associated with the hinges meeting at $\vs$, this projection introduces
a Kronecker delta into each term.  This results from projecting the
vertex-based volume associated with a hinge $\Delta V_{h \vs}$ onto a
given edge.  Hence, only those hinges meeting at $\ls$ contribute to the
edge-restricted scalar curvature;
\begin{eqnarray}
R_{\vs}  \resvol{\vs}{\ls} &=  d(d-1) \sum_{h:\; \vs, \,\ls\in h}
\frac{\epsilon_{h}}{A_{h}^{*}} \frac{1}{\binom{d}{2}} A_{h \ls \vs}A_{h}^{*}
\nonumber \\
& = 2\sum_{h:\; \vs,\, \ls\in h}
\epsilon_{h} A_{h \ls \vs} 
\end{eqnarray}
where $A_{h \ls \vs}$ is the area of the hinge $h$ restricted to both the edge $\ls$
and the vertex $\vs$~--~both $\ls$ and $\vs$ are assumed to be on $h$ otherwise $A_{h \ls \vs}=0$. 

Using this representation of the scalar curvature and the simplicial
Ricci one-form defined above, we are in position to explicitly define
the canonical form of the Einstein tensor;
\begin{eqnarray}
G_{\ls} \Delta V_{\ls} &=  R_{\ls}\Delta V_{\ls} - \frac{1}{2}\sum_{\vs\in{\ls}}
R_{\vs} \resvol{\vs}{\ls}       \nonumber \\
&= 2\sum_{h:\; {\ls}\in h} \left(\epsilon_{h} A_{h \ls} \right)   - \sum_{v\in{\ls}}
\;\;\sum_{h: \; {\ls}\in h}  \epsilon_{h} A_{h \ls \vs}        \nonumber \\
&=  2\sum_{h:\;{\ls}\in h} \left(\epsilon_{h} A_{h \ls} \right)   - 
\sum_{h:\; {\ls}\in h}  \epsilon_{h} A_{h \ls}  \nonumber \\
&=  \sum_{h:\; {\ls}\in h} \epsilon_{h} A_{h \ls}.  \label{eq:EinsteinDdim}
\end{eqnarray}

In $d=4$ this  becomes
\begin{eqnarray}
G_{\ls} \underbrace{\frac{1}{4} \ls \cdot \ls^{*}}_{\Delta V_{\ls}} &=&  \sum_{h:\;{\ls}\in h} \epsilon_{h} A_{h \ls}
\nonumber \\ 
&= &  \sum_{h:\;\ls\in h} \epsilon_{h} \underbrace{\frac{1}{2} \ls \cdot \frac{1}{2} \ls
\cot{(\theta_{h \ls})}}_{A_{h \ls \vs}}     \nonumber \\ 
G_{\ls} \ls^{*} &=& \sum_{h:\; {\ls}\in h} \epsilon_{h} \ls \cot{(\theta_{h \ls})}           
\end{eqnarray}
where $\theta_{h \ls}$ is the angle on the hinge $h$ opposite
$\ls$. Staying in $d=4$ we can check this result with the result
obtained from varying the Regge action.  In the continuum the
integrated Einstein tensor is obtained from the variational principle;
\begin{equation}
\int \sqrt{-g}\; G_{\alpha\beta} \; d^{4}x = \kappa
\frac{\delta I_{\rm geom}}{\delta g^{\alpha\beta}}
\end{equation}
where $\kappa = 16\pi Gc^{-4}$ and $I_{\rm geom}$ is the Einstein-Hilbert action. 

In RC, with action given by $\frac{2}{\kappa}\sum_{h}\epsilon_{h}A_{h}$, this becomes
\begin{equation}
G_{\ls}\ls^{*} = \kappa \frac{\delta I_{\rm Regge}}{\delta \ls}.
\end{equation}
Regge showed that the variation of the deficit angle $\epsilon_{h}$ in
the Regge action does not contribute to the final equations of motion.
Only variation of the hinge volume contributes.  Using this result we
obtain the standard Regge equations for an edge;
\begin{eqnarray}
\frac{\delta I_{\rm Regge}}{\delta \ls} &=& 2\frac{1}{\kappa }\sum_{h:\; {\ls}\in h} \epsilon_{h}
\frac{1}{2} \ls \cot{(\theta_{\ls h})}  \nonumber \\
&=& \frac{1}{\kappa} \sum_{h:\; {\ls}\in h} \epsilon_{h}\ls \cot{(\theta_{\ls h})}.
\end{eqnarray}
The integrated Einstein tensor from the variational principle is thus
found to match the result obtained from the Regge version of the
canonical Einstein tensor definition;
\begin{equation}\label{eq:Einstein4d}
G_{\ls \ls} \ls^{*} =  \sum_{h:\;  {\ls}\in h} \epsilon_{h}\ls \cot{(\theta_{\ls h})}.
\end{equation}
This agrees with the results from the moment of rotation three-form
derivations \cite{Miller:BBP, McDonald:BBP}. The factor of 2 that
explicitly appears in Eq.~(\ref{eq:Einstein4d}) that cancels the
$\frac{1}{2}$ factor in the moment arm is due to combinatoric factors
coming from the symmetry in the moment of rotation, i.e. $${\bf d}{\cal P} \wedge {\bf
  R}= {\bf R} \wedge {\bf d}{\cal P}.$$  In particular, the
integrated moment of rotation assigned to an edge is not dependent on
the ordering of the wedge product of the moment arm with the curvature
and gives rise to this numerical factor.

It is particularly instructive to confirm this result by way of
Eq.~(\ref{eq:DDRiemann}). Since the first dual acts on the space of
values, we only need note that one component survives while the second
component of the bivector contributes to the trace.  Acting on the
two-form components is the fundamental volume form, $\epsilon^{ablj}$,
which acts as a given $4$-volume.  Choosing a given component of
$G_{i}^{\phantom{j}}$ is akin to choosing an edge $\ls$ on the
simplicial lattice.  Since $R_{h}=R_{h^{*}}$ takes non-zero components
only in the directions orthogonal to hinges, the trace is the sum over
directions orthogonal to $\ls$ and $h^{*}$;
\begin{eqnarray}
\left( G^{j}, \ls^{j}\right)&=  \frac{1}{2} \sum_{h: \ls\in
  h} R_{h^{*}} \Delta V_{h^{*}\ls}  \nonumber \\
&= \frac{1}{2} \sum_{h : \ls\in h} d(d-1) \frac{\epsilon_{h}}{A_{h^{*}}}
\frac{1}{\binom{d}{2}}  A_{h \ls} A_{h^{*}} \nonumber \\
G_{\ls} \Delta V_{\ls}&= \frac{1}{4} \sum_{h: \ls\in h} \epsilon_{h} \ls^{2} \cot{(\theta_{\ls h})}
\end{eqnarray}
where we have used $A_{h \ls}= \frac{1}{2} \ls^{2} \cot{(\theta_{\ls h})}$.
Doing the usual trick of decomposing the volume on the LHS and
dividing by $\ls$, we have
\begin{equation}
G_{\ls} \ls^{*} = \sum_{h: \ls\in h} \epsilon_{h} \ls\cot{(\theta_{\ls h})}
\end{equation}
as before.   In general, the Einstein tensor in arbitrary dimension is
given by;
\begin{equation}
G_{\ls}\ls^{*} = \frac{d}{\ls} \sum_{h} \epsilon_{h}A_{h \ls}
\end{equation}
 in agreement with Eq.~(\ref{eq:EinsteinDdim}).  We thus have multiple
 methodologies for deriving the Einstein tensor, and we have shown
 that the Einstein tensor is the sum of restricted areas of  hinges
 times their associated deficit angles.

\section{Conclusion}

We have presented here the first geometric discretization of the $Ric$
in RC in arbitrary dimension.  The tracing of the Riemann tensor over
loops of parallel transport produces a one-form in the dual lattice.
Moreover, we are able to use the isomorphism between forms on the dual
with forms on the simplicial lattice to construct a simplicial
counterpart to the dual lattice Ricci one-form.  Both formulations
provide explicit meaning to the simplicial analog of the trace of the
Riemann tensor as an edge-based ``weighted average'' of curvature.  In
the dual representation the $Ric$ is a volume-weighted average while
in the simplicial representation it becomes a ratio of area-weighted
averages.

The $Ric$ defined as one-form in the simplicial or dual lattices is one
step towards accurately embedding the machinery of RF into the
piecewise-flat discretization of RC.  By representing the $Ric$, and
eventually RF, in the RC framework, we expect to be able to use RF on
geometries of arbitrary topology in arbitrary dimension.    In
particular, the 3-dimensional $Ric$ carries the full information about
the curvature of the manifold and can be used for manifold comparison
using techniques developed by Perelman \cite{Perelman:Entropy,
  Perelman:Surgery, Perelman:Time, Grove:CompGeom}.  Ongoing future work will
develop the RF equations and apply them to discrete manifolds in
higher dimension.

The definition of a $Ric$ in arbitrary dimension has further
allowed us to provide a third and independent derivation of the
Einstein tensor in RC.  By using our simplicial Ricci
one-form and the recent definition of the vertex-based scalar
curvature, we are able to write an explicit expression for
the trace-reversed $Ric$ in terms of restricted
volumes in the simplicial lattice.  This shows further the utility of
the inherent Voronoi-Delaunay duality and the associated hybrid
cells as natural volumes in RC.

\ack
  We would like to thank Shing-Tung Yau and Xianfeng Gu for
  stimulating our interest in this topic and pointing out useful
  references.  JRM would like to acknowledge partial support from the
  SFB/TR7 ``Gravitational Wave Astronomy'' grant funded by the German
  Research Foundation (DFG) and is currently supported through a
  National Research Council Research Associateship Award at AFRL
  Information Directorate.  WAM acknowledges partial support
  from the Information Directorate at Air Force Research Laboratory.
  PMA wishes to acknowledge the support of the Air Force Office of
  Scientific Research (AFOSR) for this work.  Any opinions, findings,
  and conclusions or recommendations expressed in this material are
  those of the authors and do not necessarily reflect the views of AFRL.

\appendix

\section{ Integral Volumes in Regge Calculus} \label{App:Volumes}

The canonical volumes of RC are the simplicial blocks of the lattice.
These domains define the locally flat subspaces of the geometry.  The
simplicial blocks also supply the lattice with an intrinsice
definition of local tangent spaces on which we explicitly define
vectors, tensors, and differential forms.  It is useful to decompose
these simplicial domains to fit with the character of the geometric
objects we construct.  Since all embeddings of geometric variables are
essentially integrated quantities, as opposed to the point-based
representation in the continuum, we wish the integral volumes to
reflect the nature of the object itself.  Here we provide a short
review of the methods for constructing integral volumes used in this
manuscript.

We begin by defining the simplicial volume via the inner-product of
forms.  The volume of a simplicial cell is given by the inner product
of the simplicial $d$-form with itself;
\begin{eqnarray}\label{eq:dFormIP}
\left( \polys{d}, \polys{d} \right) &=\int \polys{d} \wedge *\polys{d} \nonumber \\
&= \frac{1}{\binom{d}{d}} \norm{\polys{d}} \cdot \norm{*\polys{d}} = \norm{\polys{d}}
\end{eqnarray}
where we use the usual notation, $\norm{\cdot}$, to indicate the norm.
Since $*\polys{d}$ is a vertex of the dual lattice, i.e. the
circumcenter of $\polys{d}$, it contributes only a scalar constant to
the integral.  To ensure that the integral yields the appropriate
$d$-volume, we choose assign to any vertex a volume with unit
normalization.  Likewise, a polytope $\polyd{d}$ dual to a vertex
$\vs$ in the simplicial lattice is given by
\begin{equation}
\left( *\polyd{d},  *\polyd{d}\right) = 
\frac{1}{\binom{d}{0}} \norm{*\polyd{d}}\cdot\norm{\polyd{d}} =
\norm{\polyd{d}}
\end{equation}
where again we have $\norm{*\polyd{d}} = \norm{\vs} =1$.  Explicitly,
this volume is constructed by building local domains interior to each
simplex in the star of the vertex $\vs$ dual to $\polyd{d}$.  Using
the Voronoi construction, this volume localized on a simplex is the
set points in $polys{d}$ closest to $\vs$ than any other vertex in the
simplex.  This portion of the simplex will be called the restriction
of the simplex to $\vs$, $\resvol{\polys{d}}{\vs} = \norm{\polys{d}}_{\vs}$.
Summing over each simplex in the star of $\vs$, ${\rm St}(\vs)$, gives the
complete dual volume.
\begin{equation}
\norm{*v} = \sum_{\polys{d} \in {\rm St}(\vs)} \norm{\polys{d}}_{\vs}.
\end{equation}
We can construct arbitrary volumes that are hybrid Delaunay-Voronoi
cells through inner-products of the simplicial (dual) $r$-forms with
themselves;
\begin{eqnarray}
\left( \polys{r},\polys{r}\right) &= \int \polys{r} \wedge *\polys{r}
\nonumber \\
& = \frac{1}{\binom{d}{r}}\norm{\polys{r}} \norm{*s{r}}.
\end{eqnarray}
The factorization given by the last equatlity is a direct result of
the inherent orthogonality between the Voronoi and Delaunay lattices.
This canonical factorization is one of many factorizations.  One may
also decompose the volume associated with a given simplicial or dual
element into volumes determined by $m$-forms ($m<r$) contained in a
given $s^{(r)}$ or $n$-forms ($n>r$) in ${\rm
  St}(\polys{r})$;
\begin{eqnarray}
\Delta V_{\polys{r}} &= \sum_{\polys^{m} \in \polys{r}} \frac{1}{\binom{d}{m}}
\norm{\polys{m}} \norm{*\polys{m}}_{\polys{r}} \quad \quad &({\rm for\,
}r > m) \\ 
\Delta V_{\polys{r}} &=  \sum _{\polys{n}\in {\rm St}(\polys{r})} \frac{1}{\binom{d}{n}}
\norm{\polys{n}}_{\polys{r}}  \norm{*\polys{n}} \quad  &({\rm for\,
}r  < n).
\end{eqnarray}
Here, the Voronoi-Delaunay duality is again particularly useful as it
allows us to construct the restricted volume via restriction of only a
subspace of a given volume.  The restriction is applied to the
subspace such that the restriction makes sense, i.e. the restriction of
$\polys{m}$ to $\polys{r}$ ($r>m$) trivially yields the norm  $\norm{\polys{m}}$.  Such restrictions are explicitly used in
the definition of the vertex-based scalar curvature which require
vertex $d$-volumes to be decomposed using the vertex-restriction of the
hinge area \cite{McDonald:RCscalcurv}

\section{ Operations on Discrete Forms} \label{App:DisFormOps}

In the lattice we endow the geometry with two distinct spaces of
differential forms, (1) the simplicial skeleton as the representation
of the homology and (2) the dual skeleton as the
representation of the cohomology.  The representation
of differential forms on a simplicial complex is based on the ideas of
Whitney \cite{Whitney:book} and has been used in computational
electromagnetism \cite{Bossavit:1991, Bossavit:1998, Arnold:2006} and computational geometry
\cite{Gu:2002, Gu:2003}.   The purpose of such a representation is to
not just discretize tensor and differential form fields by
representing their components point-wise on some discrete set of
points, but to embed the full geometric character of a field in the
discretization.   In this way, one hopes to preserve the general
geometric properties and symmetries of the field in the
discretization.    In this appendix we review some useful
isomorphisms between the spaces of forms in the simplicial and dual
lattices.

The first and most straightforward isomorphism is the Hodge dual.  The
Hodge dual maps an element of $\Lambda^{(r)}$ ($\Lambda^{*\, (r)}$) to
$\Lambda^{* \, (d-r)}$ ($\Lambda^{(d-r)}$).  This is  defined by
mapping the scalar weighting to a given simplicial (dual) element of
the skeleton to its geometric dual, i.e.
\begin{equation}
\alpha_{s^{(r)}} \longrightarrow \alpha_{*s^{(r)}}.
\end{equation} 
This is done via formal mapping \cite{Desbrun:DisForms}
\begin{equation}
\frac{1}{|s^{(r)}|} \left< \alpha, s^{(r)}\right> =
\frac{1}{|*s^{(r)}|} \left< *\alpha, *s^{(r)}\right> 
\end{equation}
where $\left< \alpha , \Omega\right> = \int_{\Omega} \alpha$. Since
differential forms in RC are represented as scalar weights
on elements of the lattice, this isomorphism is a simple mapping of
the weight from an element on one lattice to its dual element.

We also can construct the raising (lowering) operations in the
lattice.  In the continuum, this operation is carried out via the
metric or its inverse applied to components of the form.  In the
lattice, we must construct a way of identifing a scalar weighting to an
$r$-form of the simplicial (dual) lattice using the weights of the
$r$-forms in the dual (simplicial) lattice.  We define the isomorphism
taking dual $r$-forms to a simplicial $r$-form as
\begin{equation}
\alpha_{\polys{r}} \Delta V_{\polys{r}} =\left\{\begin{array}{cc} \displaystyle{
    \sum_{\polyd^{r}:\,\,  \polys{r} \in *\polyd^{r}}
    \alpha_{\polyd{r}}}  \resvol{\polyd{r}}{\polys{r}},
  &\mathrm{if\; }2r\leq d \\
 \phantom{as} &\\
\displaystyle{ \sum_{\polyd{r}:\,\,  *\polyd{r}\in \polys{r}}}
\alpha_{\polyd{r}} \resvol{\polyd{r}}{\polys{r}}, &\mathrm{if\;
}2r>d\end{array} \right.
\end{equation}

Using the orthogonal decomposition and restriction of volumes defined
in Appendix ~\ref{App:Volumes}, the volumes on the RHS are given by
\begin{equation}
\resvol{\polyd{r}}{\polys{r}} =\left\{\begin{array}{cc} \frac{1}{\binom{d}{r}}
 \norm{\polyd{r}}  \norm{*\polyd{r}}_{\polys{r}},  & \mathrm{if\;
 }2r\leq d \\
 \phantom{as} & \\
\frac{1}{\binom{d}{r}} \norm{*\polyd{r}}
\norm{\polyd{r}}_{\polys{r}}, &\mathrm{if\; }2r> d 
\end{array} \right.
\end{equation}
One can define a similar isomorphism from the simplicial lattice to
the dual lattice by taking the sum over elements of the simplicial
skeleton.  It is important here that we incorporate the restriction of
the integral domain into the defintion to ensure that if we apply the
inverse isomorphism that we reobtain the initial $r$-form.  This can
be easily checked.
\vspace{.25in}

\bibliography{Ricci}{}
\bibliographystyle{unsrt}

\end{document}